\documentclass[onecolumn]{aa} 
\usepackage{graphicx}
\def\beq{\begin{equation}}
\def\eeq{\end{equation}}
\def\bey{\begin{eqnarray}}
\def\eey{\end{eqnarray}}

\def\kpc{\, {\rm kpc} }

\def\sun{\odot}

\def\lsun{L_\odot}

\def\vg{{\bf g}}
\def\vR{{\bf R}}
\def\grad{{\bf \nabla}}
\def\div{{\bf\nabla\cdot}}
\def\a0{$a_0$}

\begin{document}
\title{Roche Lobe Shapes for testing MOND-like Modified Gravities}
\author{HongSheng Zhao\inst{1}\inst{2}
\thanks{Outstanding Young Overseas Scholar and PPARC Advanced Fellow} \and
LanLan Tian\inst{3}\inst{4}}
\institute{National Astronomical Observatories, Chinese Academy of
Sciences, Beijing 100012, PRC 
\and
SUPA, School of Physics and Astronomy, University of St Andrews, KY16 9SS, Fife, UK,
\and 
Purple Mountain Observatory, Chinese Academy of
Sciences, Nanjing 210008, PRC 
\and
University of Victoria, 
Department of Physics and Astronomy, British Columbia, V8P 5C2, Canada}
\offprints{H. Zhao, \email{hz4@st-and.ac.uk}}
\date{}
\abstract{  
Dark Matter (DM) theories and mass-tracing-light theories like MOND are by construction nearly degenerate 
on galactic scales, but not when it comes to the predicted shapes of Roche Lobes of a two-body system 
(e.g., a globular cluster orbiting a host galaxy).  We show that 
the flattening of the Roche lobe is sensitive 
to the function $\mu(g)$ in modification of the law of gravity.
We generalise the analytical results obtained in the deep-MOND limit by Zhao 
(2005, astro-ph/0511713), and consider a binary in the framework of 
a MOND-like gravity modification function $\mu(g)$ or
a general non-Keplerian gravity $g \propto R^{-\zeta}$.   
We give analytical expressions for the
inner Lagrange point and Robe lobe axis ratios.  The Roche lobe volume
is proven to scale linearly with the true mass ratio, which applies
to any $\mu(g)$, hence mass-tracing light models would overpredict 
the Roche lobe of a DM-poor globular cluster in a DM-rich host galaxy, 
and underpredict the size of a DM-richer dwarf satellite.  
The lobes are squashed with the flattening $\sim 0.4$ 
in the strong gravity and $\sim 0.6$ in the weak gravity; a precise measurement 
of the flattening could be used to verify the anisotropic dilation effect 
which is generic to MOND-like gravity.  
We generalise these results for extended mass distribution, and compare
predicted Roche radii with limiting radii of 
observed globular clusters and dwarf galaxy satellites.  
 }
\maketitle
\section{Introduction}

The law of gravity is still uncertain experimentally on very small scales and very
large scales.  An example is the running controversy (Sellwood \& Kowsowsky 2001, 2002 and references therein) 
over dark matter or MOND (Modified Newtonian Dynamics) 
as the right explanation of why the Newtonian gravity $g_N$ from baryonic stars and gas
in galaxies fall short of explaining the observed acceleration $g$ by a factor
$\mu(g)=\left({g^n \over a_0^n+g^n}\right)^{1 \over n} \sim (0.01-1)$,
where $a_0 \sim 10^{-8} {\rm cm}\sec^{-2}$ is a characteristic accerleration,
and
$n=2$ (see examples in Sanders \& McGaugh 2002) or $n=1$ (Zhao \& Famaey 2006).
This is normally the
justification to invoke dark matter particles on galaxy scales so that $M/L \sim
1/\mu \sim (1-100)$;
the mass profile of the dark particles needs to be well-coupled to the baryon distribution (McGaugh 2005). 
Alternatively  often a good fit can be made in a MOND gravity (Milgrom 1983) 
or MOND-like gravities (Bekenstein 2004, Sanders 2006).
The important question for astronomers 
is whether these two views of fundamental difference are by and large 
equally good {\it descriptions of features} in galactic systems 
(e.g., the orbits of halo streams in the Milky Way, Read \& Moore 2005), hence are degenerate,
or there are structures in galaxies which simply cannot be explained by 
any modifications of the law of gravity (e.g., problems on sub-galactic scale, Zhao 2005).

In MOND the gravity $\vg = -\grad \Phi$ 
and gravitational potential $\Phi$ follow a Poisson-like equation
\beq\label{binary}
- \div { \vg_\mu \over 4\pi G} = -\div {\vg_N \over 4\pi G} = 
\sum_{i=1}^N M_i \delta\left(\vR_i\right), ~\vg_\mu \equiv \vg \, \mu(|\vg|),
\eeq
for an isolated system of $N$ baryonic point masses $M_1$, ..., $M_N$ at distances $R_i$.
In general, the vector $\vg_\mu$ is not curl-free, and equals the 
curl-free Newtonian gravity $\vg_N$ plus a divergence-free curl-field.
Here $0< \mu(g) \le 1$ is the modification factor, and is a function of the amplitude
of the gravity $g=|\vg|$ (or equivalently an implicit function of $g_\mu=|\vg_\mu|$).
if $\mu=1$, then $\vg=\vg_\mu=\vg_N$ and the normal Poisson's equation is recovered.
This modification satisfies the usual conservations of total energy and angular momentum
for any geometry of baryonic distribution (see Bekenstein \& Milgrom 1984).  

In the MOND theory and in spherical symmetry, the gravity
$g(R)$ at distance $R$ from a baryonic point mass $M$ is boosted from
the Newtonian value $g_N(R) = GM/R^2$ by a factor $1/\mu>1$ such that
far away from the mass point we have
\beq
\mu = {g_\mu \over g} \sim {g_N/a_0 \over g/a_0} \sim \left({g \over
a_0}\right)^1 \sim  \left({R \over R_0}\right)^{-1},  \eeq 
This reproduces the flat
rotation curves in bright disk galaxies
at radii beyond $R_0=\sqrt{GM_0 \over a_0}$ since $V^2 \sim g R \sim a_0 R_0 =cst$.
The predictive power of this 20-year-old classical theory with
virtually no free parameters (Bekenstein \& Milgrom 1984) is recently
highlighted by the astonishingly good fits to contemporary kinematic
data of a wide variety of high and low surface brightness spiral and
elliptical galaxies; even the fine details of the ups and downs of
velocity curves are elegantly reproduced without fine tuning of the
baryonic model (Sanders \& McGaugh 2002, Milgrom \& Sanders 2003).
For a long time a problem has been that the gravitational lensing 
in this non-relativistic theory is illly-posed, and 
it misses a factor of two in the bending angle (e.g., Qin, Wu \& Zou 1995).
However, this empirical MOND has now a respectable relativistic
field theory formulation (called TeVeS by Bekenstein 2004), which
passes standard tests to check General Relativity, and allows for
rigourous modeling of Hubble expansion and gravitational lensing.
This has generated wide interests, and many are examining the consequences
of modifications to gravity (Skordis, Mota, Ferreira et al. 2005, 
Hao \& Akhoury 2005, Chiu, Ko \& Tian 2005, Zhao, Bacon, Taylor et al. 2006,
Pointecouteau \& Silk 2005, Ciotti \& Binney 2004, Ciotti, Londrillo, Nipoti, 2006, 
Baumgardt, Grebel, Kroupa 2005, Read \& Moore 2005,
Famaey \& Binney 2005, Zhao \& Famaey 2006).  

Zhao (2005) has shown that the shape of the Roche lobes in the deep-MOND limit
is more squashed than in the Newtonian case.  
Here we study the shape of the Roche lobe of a two-body system in a MOND-like gravity.  
We show that Roche lobes shape varies with the assumed law of gravity, and 
could be used to differentiate among laws of gravity.  The results are first 
shown for point masses and are generalised to extended mass distribution.

\section{MOND-like gravity}

\begin{figure}{}
\resizebox{9cm}{!}{\includegraphics{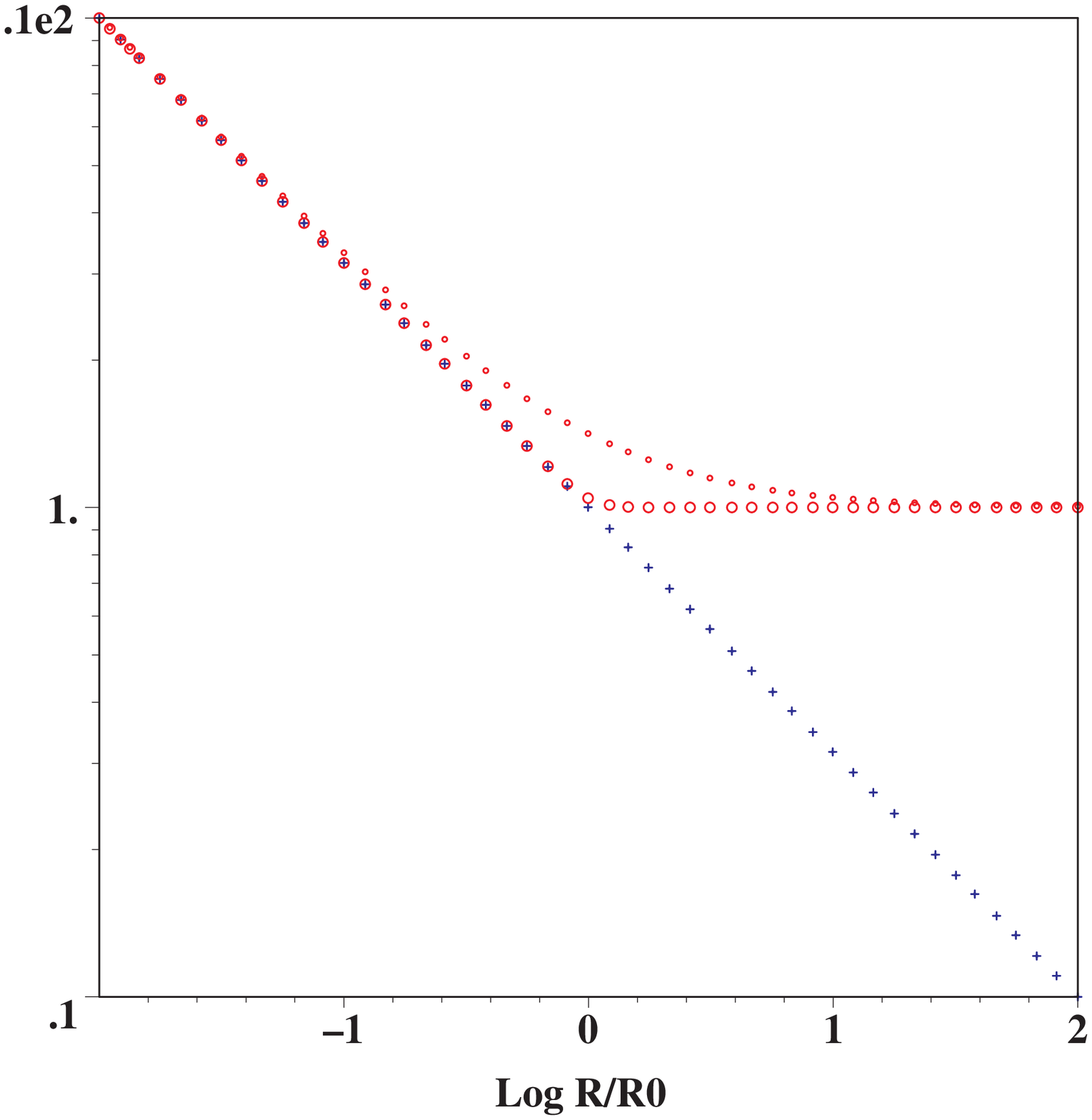},\includegraphics{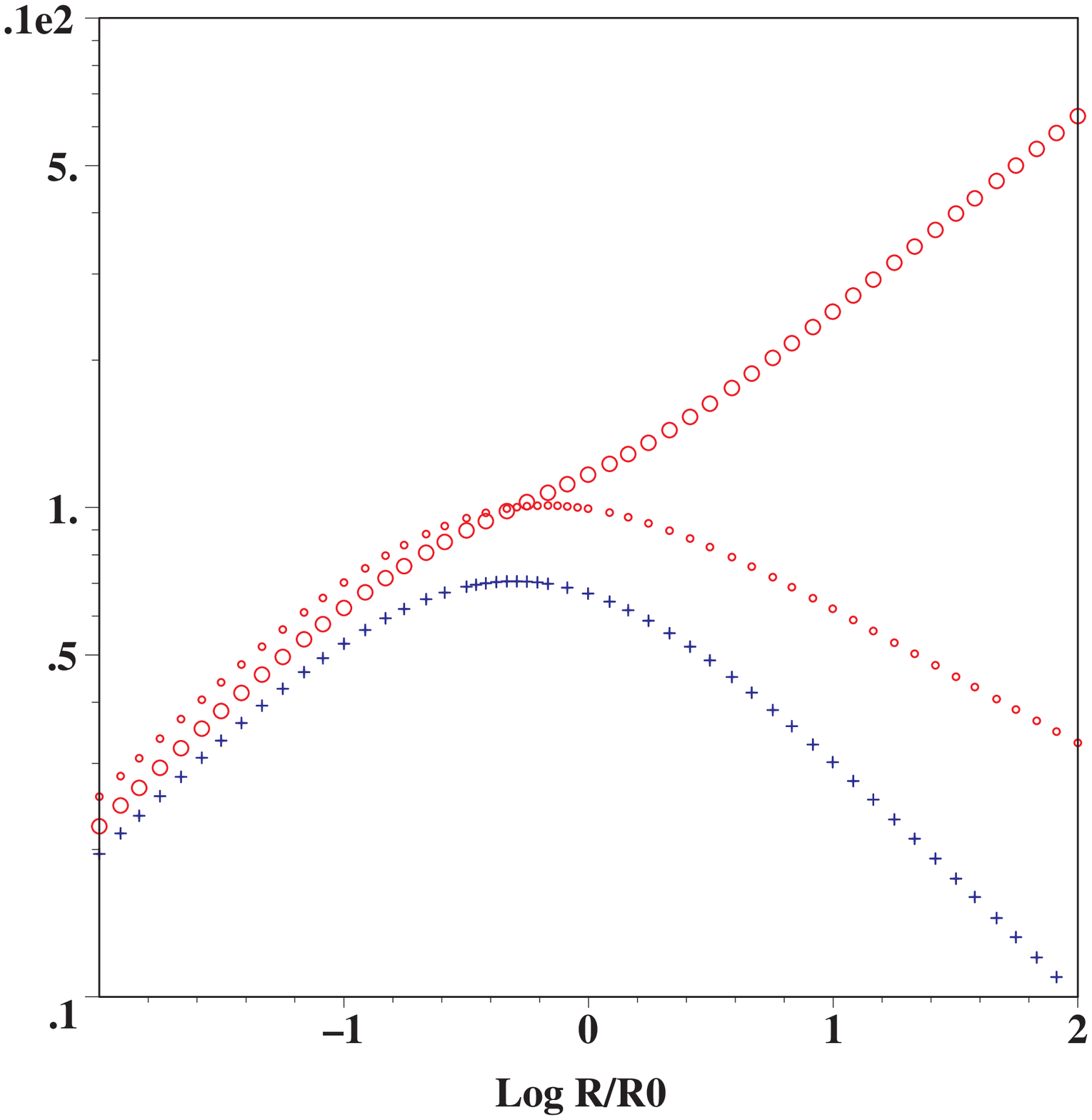}} 
\vskip 0.5cm
\resizebox{9cm}{!}{\includegraphics{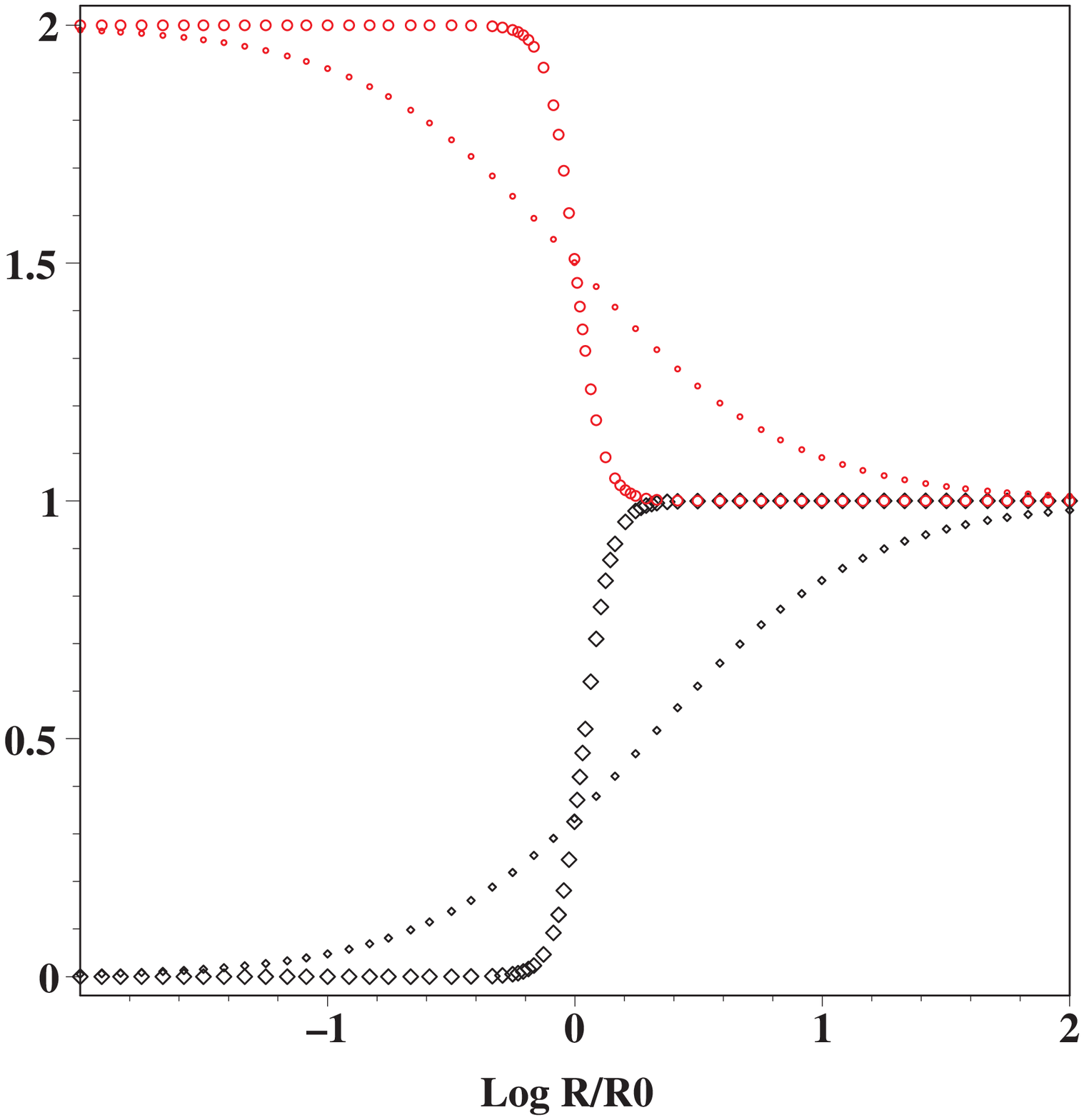},\includegraphics{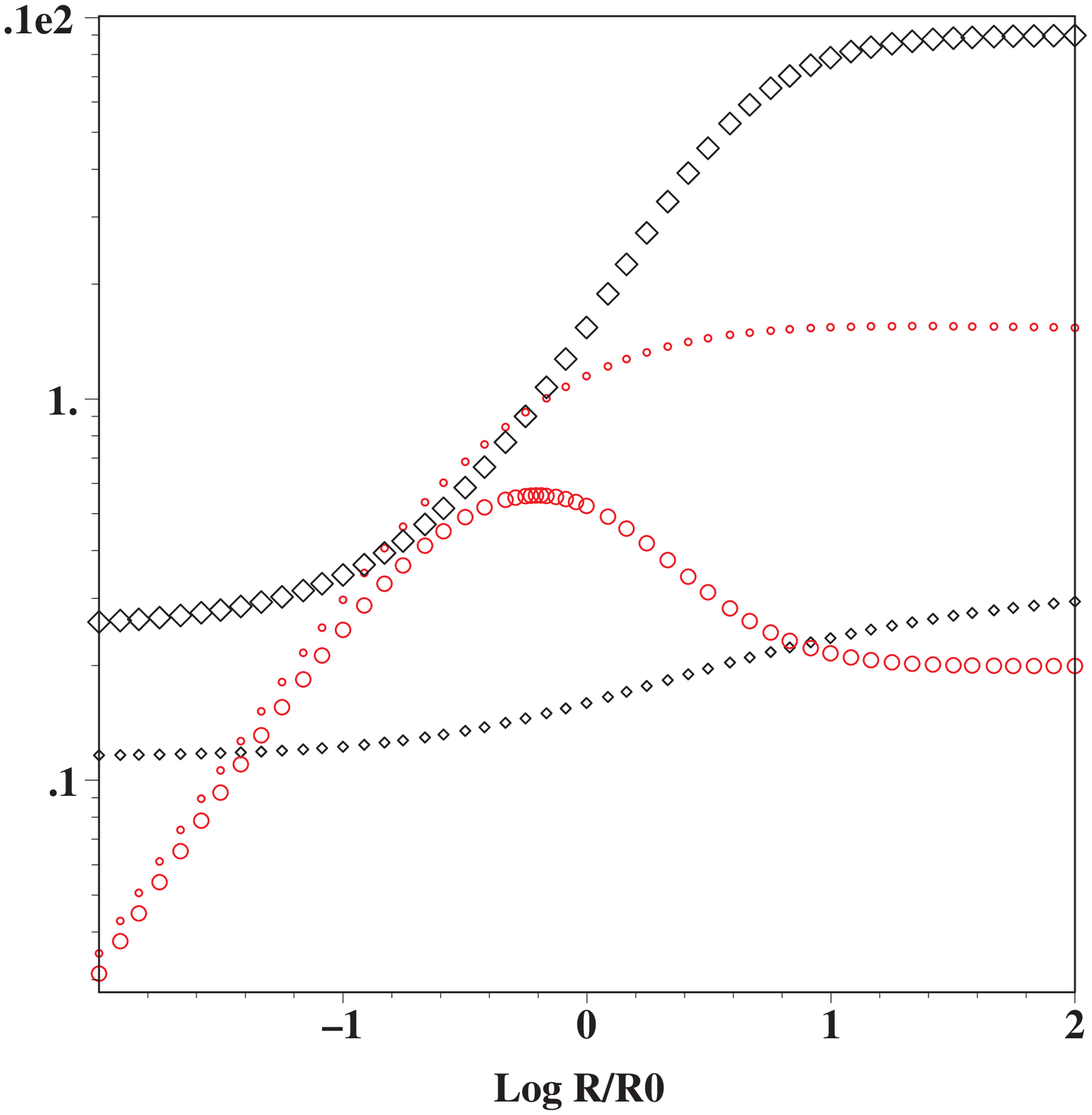}} 
\vskip 0.5cm
\resizebox{8cm}{5cm}{\includegraphics{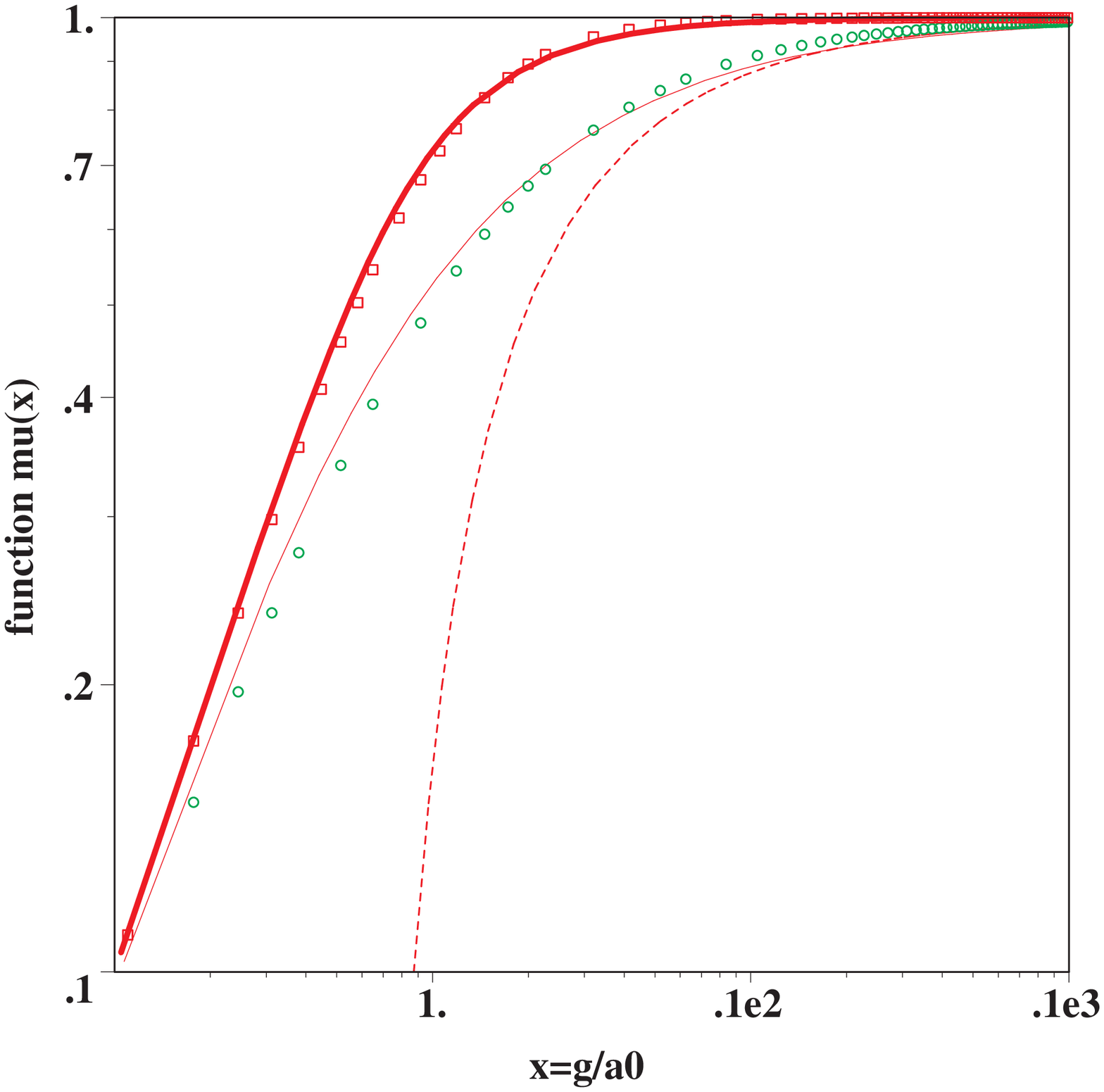}} 
\caption{
{\bf Top panels}: shows 
the circular velocity curves $V_{\rm cir}(R)=\sqrt{g R}$ in various modified gravity (red circle)
(cf. eq.~\protect{\ref{mu}}) 
and $V_{\rm cir}(R)=\sqrt{g_N R}$ in Newtonian gravity (blue cross); both in units of $\left(GM_0a_0\right)^{1/4}$.
These are shown in {\bf left panels} for point mass with
$(k,n)=(1/2,1)$ (smallest symbols), $(k,n)=(1/2,8)$ (bigger symbols)
and in {\bf right panels} for an extended Hernquist profile mass with a scale $b=1/2$ in 
a stronger-than-MOND gravity with $(k,n)=(9/10,1)$.
{\bf Middle panels}: Same parameters as in upper panels, 
except showing $\zeta$ (red circles) and $\Delta$ (black diamonds) 
as functions of rescaled orbital distance $R/R_0$. 
{\bf Bottom panel}: compares the function $\mu(g)$ in the popular MOND models ($\mu(x)=x/\sqrt{1+x^2}$ in boxes, and
$x/(1+x)$ in circles) with our function in the case $(k,n)=(1/2,3)$ (thick line) and $(1/2,3/2)$ (thin line)
and $(1,1)$ (dashed line).  
}\label{zdr}
\end{figure}

A very broad class of modified gravity models could be parametrized by a 
modification function given as follows,
\beq\label{mu}
{1 \over \mu} \equiv {g \over g_\mu} = \left[ 1+\left( {a_0 \over g_\mu}\right)^{k n}\right]^{1 \over n},
\eeq
which makes $\mu(g)$ an implicit function of the gravitational strength $g=|{\bf g}|$.
The Newtonian gravity corresponds to models with $k=n=0$.  Exotic gravity can be achieved 
by letting, e.g., $(k,n)=(1,1)$
\footnote{so that the gravity $g \rightarrow cst$ at large distances as in, e.g., (Mannheim 1997).}
or $(k,n)=(-{3 \over 2},1)$\footnote{so that $g \propto R^{-5}$ at small distances as in, e.g., the theory of 
Qin et al. 2005 with three extra nanometer-thick spatial dimensions.}.  Example rotation curves are shown 
in Fig.~\ref{zdr} for a point mass (upper left panel) and an extended mass (upper right panel).
Conventional MOND gravity corresponds to models with $k=1/2$.  
Such a model in spherical symmetry goes from $g \sim g_\mu \sim g_N = GM R^{-2}$ in strong gravity to 
$g \sim \sqrt{g_Na_0} = \sqrt{GMa_0} R^{-1}$ in weak gravity around a point mass $M$.
The sharpness of the transition is controlled by the parameter $n$.  
The location of the transition zone for a point mass $M$ is given by $R_0 \equiv \sqrt{GM \over a_0}$.
To see the link with conventional MOND $\mu$ function, we note that for $k=1/2$ equation~(\ref{mu}) can be
inverted as 
\beq
\mu(x) = x \left[ {1 \over 2} + \sqrt{{1 \over 4} + x^{n}} \right]^{-2/n}\!\!\!\!, 
~x={g \over a_0}, ~{\rm for~} k={1 \over 2},
\eeq
which approaches $x$ or $1$ for small or big $x$.
Our function in the case $(k,n)=(1/2,3)$ approximates the standard MOND function $\mu=x (1+x^2)^{-1/2}$ very well
(cf. bottom panel of Fig.~\ref{zdr}).
And the case with $(k,n)=(1/2,3/2)$ approximates the simple MOND function $\mu=x (1+x)^{-1}$ recommended by 
Zhao \& Famaey (2006).

For more compact notations later on we shall introduce a few auxilary functions, related to $\mu$
in a spherical potential.  We define
\beq
\Delta \left(g \right) \equiv {d\ln \mu \over d\ln g}, \qquad \Delta_1 \equiv \Delta+1={d\ln g_\mu \over d\ln g}.
\eeq
We also define
\beq
\zeta (g) \equiv - \left( { d \ln R(g) \over d \ln g } \right)^{-1}\!\!\!,\qquad  
\zeta_1 \equiv \zeta + 1 = -{d\ln \Omega^2 \over d\ln R},
\eeq
where $\Omega(R)=\sqrt{g \over R}$ is the angular frequency for a circular orbit of radius $R$.
In writing down $R(g)$ we treat the spherical radius $R$
as a function of the gravitational field strength $g$ through
the relation $M R^{-2} \propto  g_N \sim g_\mu = \mu g $.
Note in the case of the field around a point mass, and assume non-negative $n$ and $k$, 
$\Delta(g)$ is related to the parameter $\zeta(g)$ by
\beq
{\zeta \over 2} = {d\ln g \over d\ln R^{-2}} = {1 \over \Delta_1} =\cases{1 & near a point mass  \cr
                                             1-k & far from a point mass.  \cr}
\eeq
Conventional MOND models have the property that 
$(\Delta, \zeta) \rightarrow (0,2)$ when the gravity is strong and
$(\Delta, \zeta) \rightarrow (1,1)$ when the gravity is much weaker than $a_0$.
The result in this paper applies to any function $\Delta(g)$ or $\zeta(g)$. 
In general, 
we shall call the above gravity $g$ the MOND-like Gravity or Modified Gravity.

Fig.~\ref{zdr} shows how $\zeta$ (red circles) and $\Delta$ (black diamonds) 
change as the MONDian gravity decreases away from a point mass 
(middle left panel) and from an extended mass (middle right panel).
These MOND models have a narrow range of $1 \le \zeta \le 2$ or $1 \ge \Delta \ge 0$.

\section{Roche Lobes of a two-body system in a MOND-like gravity}

A binary system is specified by two baryonic masses $M$ and $m$ separated by a distance $D_o$ along, say, 
the z-axis.  The two bodies could be a galaxy and a satellite.
The MOND-like gravity distribution is determined by equation~(\ref{binary}) where $N=2$ for a binary. 
Define the inner Lagrange point of the binary at a distance $r_1$ and $D_o-r_1$ from the masses $m$ and $M$.
There are only 3 independent dimensionless quantities in this system, 
$r_1/(D_o-r_1)$, $m/M$ and a typical value for the parameter $\zeta$.  
From dimensional arguments we expect that the Roche radius might be expressable in general as
\beq
{r_1 \over D_o} = \left({m \over A(\zeta) M}\right)^{B(\zeta)}, 
\eeq
where $A$ and $B$ are dimensionless functions with dimensionless arguments; 
we now go on to show, using perturbation analysis, that this is indeed the case, and we set 
the exact expressions for $A$ and $B$.

We set up the coordinate system such that the origin is on the smaller mass $m$, 
at a distance $D_o$ from the mass $M$ on the z-axis.  
Let the low-mass satellite with $m/M \ll 1$ rotate around the mass $M$ (fixed) with an angular velocity 
$\Omega \hat{\bf y}$, 
then particles in the corotating frame conserve  the Jacobi energy with an effective (triaxial) potential
\beq\label{omega}
\Phi_e(x,y,z) \equiv \Phi - {x^2 + (z-D_o)^2 \over 2}\Omega^2, ~\Phi \equiv \Phi_0 + \Phi_1,
\eeq
where the gravitational potential $\Phi$ of the galaxy-satellite system is expressed 
as two terms, where we treat the effect of the smaller mass $m$ as a perturbation $\Phi_1$ 
to the unperturbed potential $\Phi_0$ of the bigger mass $M$.

First consider the unperturbed case where we set the satellite $m=0$.  
At a distance $R$ far way from the isolated baryonic mass $M$,
the gravitational potential can in general be approximated as spherical with 
\beq
\Phi_0  = \int g(R) d R, \qquad R=\sqrt{(z-D_o)^2+y^2+x^2}
\eeq
where the gravity $g(R)$ at radius $R$ from the spherical galaxy centre is related
to the Newtonian gravity $-GM{\bf R}/R^3$ by
\beq
\vg_\mu = \vg_N = - {G M \over R^3} [x,y,z-D_o], \qquad \vg=-\grad \Phi_0(R).
\eeq
Taylor-expand the three components of the gravity in the vicinity of the origin $(x,y,z)=(0,0,0)$, we have
\beq
\grad \Phi_0 \approx  [x, y, D_o] \Omega^2 + [0,0,z ] { d g(D_o) \over dD_o},
~ \Omega^2 \equiv {g(D_o) \over D_o}.
\eeq
The gravity field is such that 
if a test particle is put on circular orbit with an orbital radius $R=D_o$ around $M$,
then the rotation angular frequency must be $\Omega$ in order to 
balance gravity with centrifugal force.
Integrate the above potential gradient, we find
\beq
\Phi_0(R) \approx \Phi_0(D_o) + \left [ x^2 + y^2  - z^2 \zeta  \right] {\Omega^2 \over 2} + z D_o \Omega^2
\eeq
in the vicinity of the origin $(0,0,0)$, where according to the definitions of $\zeta$ and $\Delta$ we have
$\zeta  = -{  d\log g(D_o)  \over d\log D_o}$.

\begin{figure}{}
\resizebox{9cm}{!}{\includegraphics{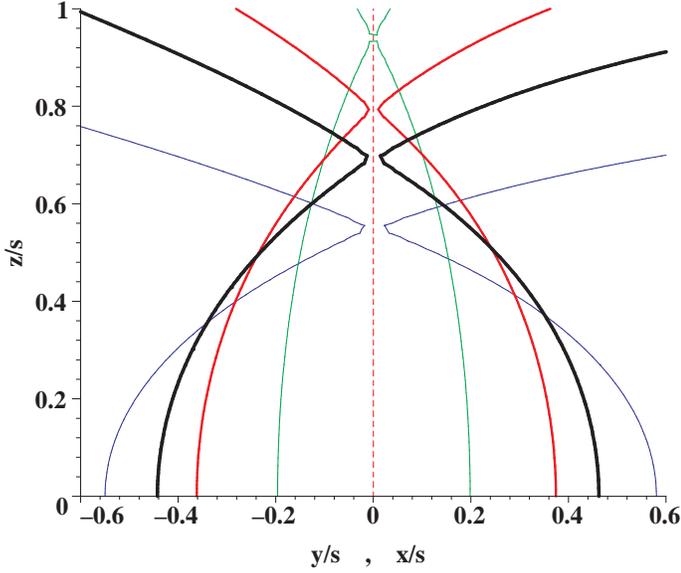}} 
\caption{
shows one octant of the 3-dimensional Roche Lobe of a satellite of mass m rotating in the xz plane around a mass M at a distance $D_o$ on the z-axis.  The sizes are scaled by a length $s \equiv \left({m \over M(D_o)}\right)^{1/3} D_o$.  
The curves are for $\zeta = -{d \ln g \over d \ln D_o}=2$ (Newtonian, thick black) and $\zeta=1$ (Deep-MOND, red).
A slimer Roche lobe for  $\zeta=1/5$ (thin green) and a fatter Roche lobe for $\zeta=5$ (thin blue) are also made 
for comparison. }\label{lobe}
\end{figure}

Now we consider the perturbation due to the satellite. 
Milgrom (1986) showed that in a medium under a uniform external field $g_0 \hat{z}$ with a dielectric
index $\mu(g_0)$, the perturbation due to the added point-mass $m$ is 
almost Newtonian apart from a mild anisotropy with a potential (cf. his eq. 11)
\beq\label{pert}
\Phi_1 (x,y,z) = - {G  m' \over \tilde{r}}, ~\tilde{r}=\sqrt{z^2 + \Delta_1\left(y^2+x^2\right)}, 
\eeq
where $m'=m/\mu(g_0)$ is the modified inertia of the point mass, 
and $\tilde{r}$ is the effective distance from the centre of the satellite and 
$\Delta_1 = \Delta +1 = 1 + {d \log \mu(g_0) \over d\log g_0} \ne 1$ is the MOND dilation factor.  
Here we show that we can recover the same expression for a satellite in a galaxy.
Sufficiently far away from the satellite, the gravity is dominated by the spatially slow-varying gravity field 
of the unperturbed galaxy, which we denote with the shorthand $g_\vR \equiv |\vg(\vR)|$.  We also use the shorthand 
$\mu_\vR \equiv \mu(|\vg(\vR)|)$ to denote the slow-varying MOND "dielectric index" due to the galaxy.
The perturbed dielectric index is $\mu_\vR$ plus a perturbation due to the self-gravity of the satellite,
which is given by
\beq
\mu(|\vg|) - \mu_\vR \approx  {d\mu \over d g} \left[|\vg|-g_\vR \right] \approx  
 {d\mu \over d g} \hat{\vR} \cdot \grad \Phi_1.
\eeq
The perturbation part of the modified Poisson's equation (eq.~\ref{binary}) becomes 
\beq
\grad \cdot {\mu_\vR \vg_\vR \over 4 \pi G}
- \grad \cdot { \left( \mu_\vR + {d\mu \over d g} \grad_\vR \Phi_1 \right) (\vg_\vR -\grad \Phi_1) \over 4 \pi G } 
\approx m \delta(r).
\eeq
Cancel out the 0th order terms due to the mass $M$ only, drop terms of order $|\grad \Phi_1|^2$ and small terms 
of order ${1 \over R} \grad \Phi_1$, 
we end up with a linear equation for $\Phi_1$, which in the vicinity of the satellite is given by
\beq
\left(\Delta_1 \partial_z^2 + \partial_x^2 + \partial_y^2 \right)\Phi_1 = 4\pi G m' \delta(r),
\eeq
where 
\beq\label{mp}
{G m' \over G m} \approx {1 \over \mu_\vR} \approx {g(D_o) \over g_\mu(D_o)} = { \Omega^2 D_o \over  GM D_o^{-2} }.
\eeq
Apply the transformation $z'=z\sqrt{\Delta_1}$, $x'=x$, $y'=y$, it can be shown that
to the first order in $m$ the perturbation in potential is given by equation~(\ref{pert}), 
where $g_0 \approx g_\vR \approx g(D_0)$.  As verified by Milgrom (1986) using a numerical solver, 
our equation (14) is "a good approximation" as long as the Newtonian force of the satellite
${G m \over r^2}$ is small compared to the Newtonian part of the external field $g_0 \mu(g_0)$,
"thus the equipotential surfaces are ellipsoids of revolution with the major axis
in the direction of the external field."

The combined potential $\Phi(x,y,z)  = \Phi_0 (R) + \Phi_1 (x,y,z) $ 
clearly has axial symmetry around the z-axis, and has two centres separated by distance $D_o$ along the z-axis.
The above formulation approximates the potential of, e.g., the Milky Way galaxy of baryonic mass $M$
plus a satellite of baryonic mass $m$.\footnote{\bf To check the validity of our approximations, we also compute 
the density $\rho$ from the modified Poisson's equation 
$\rho =-{1 \over 4 \pi G} \grad \cdot \vg_\mu$, and the mass $m(r)$ enclosed in a radius $r$ from 
${G m(r) \over r^2} = - \int {d{\bf \Omega} \over 4\pi} \cdot \vg_\mu$.  We find that the density  
$\rho \sim 10^{-10} {M \over D_0^3} \sim 0$, and 
${m(r) \over m}-1 \sim {10m \over M} \sim 0$ unless very close to the smaller pointmass $m \sim (10^{-6}-10^{-3})M$ 
or the bigger pointmass $M$.
The approximations are less good if the bigger mass $M$ is extended.  }

The effective potential $\Phi_e(x,y,z)$, if Taylor expanded near the secondary $(0,0,0)$, 
has an triaxial shape with
\beq
\Phi_e \approx cst  + \Omega^2D_o^2 \left[ {y^2 - \zeta_1 z^2 \over 2D_o^2} - {D_o m/M\over \sqrt{z^2+ \Delta_1 (y^2+x^2)}} \right], 
\eeq
where the terms to the first order of $z$ and second order to $x$ are canceled due to force balance.
The inner or outer Lagrangian points is then calculated from the saddle point of the effective potential where
\beq
0 = \left.{\partial \Phi_e(0,0,z) \over \partial z} \right|_{z=\pm r_1} 
= \Omega^2 D_o \left[ \mp  {r_1 \zeta_1 \over D_o} \pm {m \over M } {D_o^2 \over r_1^2} \right].
\eeq
This definition of the Lagrange radius $r_1$ reduces to
\beq\label{gmr}
\left({m \over r_1^3}\right)  = \zeta_1 \left({M \over D_o^3}\right).
\eeq
So inside the Lagrange radius $r_1$ the average density of the satellite equals k-times
the average density inside the orbital radius $D_o$ of the satellite.  
The equation can also be written as 
\beq\label{lag}
{r_1 \over D_o} =  \left({m \over \zeta_1 M}\right)^{1 \over 3}, 
\qquad \zeta_1=1+\zeta.
\eeq
Note that the masses here $m$ and $M$ are true baryonic masses of the binary, 
{\it not} the modified inertia masses.  Eq.~(\ref{gmr}) can be further massaged into
\beq\label{resonance}
\Omega' \equiv \sqrt{G m' \over r_1^3} = \sqrt{\zeta_1} \Omega,
\eeq
where $\Omega'$ takes the meaning of the angular frequency of circulation around the secondary at its Robe radius.
Hence the tidal radius of the secondary scales linearly with the terminal velocity of the secondary,
and the internal period is within a factor of $\sqrt{\zeta_1}$ of the period of the secondary's orbit.
Finally note that near the tidal radius, the gravity due to the secondary and primary has the ratio 
\beq
{G m' \over r_1^2 g(D_o) } = \zeta_1 \left({m \over M}\right)^{1 \over 3} \ll 1,
\eeq
which confirms that near the tidal radius $r_1$ it is valid to approximate the self-gravity of a low-mass secondary 
($m/M \ll 1$) as a perturbation to the slow-varying gravity $g(D_o)$ of the primary.  

The shape of the Roche Lobe is specified by contour of equal effective potential with the contour passing the Lagrange points 
$(0,0,\pm r_1)$.  The effective potential can be redefined as a dimensionless function $\Psi_e(x,y,z)$ 
after massaging zero point and a prefactor.  Substitute $(m/M)$ in favor of $r_1/D_o$ using eq.~(\ref{lag}), we found 
that the points $(x,y,z)$ along the Roche lobe satisfy
\beq\label{Psi}
\Psi_e = {3 \zeta_1 r_1^2  + y^2 - \zeta_1 z^2 \over 2 r_1^2} - { \zeta_1 r_1 \over \sqrt{z^2 + \Delta_1 y^2+ \Delta_1 x^2}} = 0.
\eeq
Clearly the z-axis intersects the Roche Lobe at $(0,0,\pm r_1)$, where
\beq\label{r1d}
{r_1 \over D_o} = \left({m \over \zeta_1 M}\right)^{1 \over 3},  
\eeq 
which defines the first axis of the Roche Lobe.
Draw a line parallel to the x-axis direction, it intersects the Roche lobe at $(\pm r_2,0,0)$.  Solving equation~(\ref{Psi}) 
we find
\beq\label{ratio21}
{3r_2 \over 2r_1} = {1 \over  \sqrt{\Delta_1} },
\eeq
which defines the second axis of the Roche Lobe.
Likewise the third axis of the Roche Lobe is defined by the intersection points $(0,\pm r_3,0)$
with a line parallel to the rotational y-axis.  We find
\beq\label{ratio32}
{2 r_3 \over 3 r_2}  = (u^2-1)^{1 \over 3}\left[(u+1)^{1 \over 3} - (u-1)^{1 \over 3}\right],
~ u\equiv \sqrt{1+\Delta_1\zeta_1}
\eeq
by simply solving equation~(\ref{Psi}), which reduces to an essentially cubic equation
\beq\label{cubic}
{r_2 \over r_3} - 1 = {4 \over 27 \Delta_1 \zeta_1}\left({r_3 \over r_2}\right)^2.
\eeq

\begin{figure}{}
\resizebox{9cm}{!}{\includegraphics{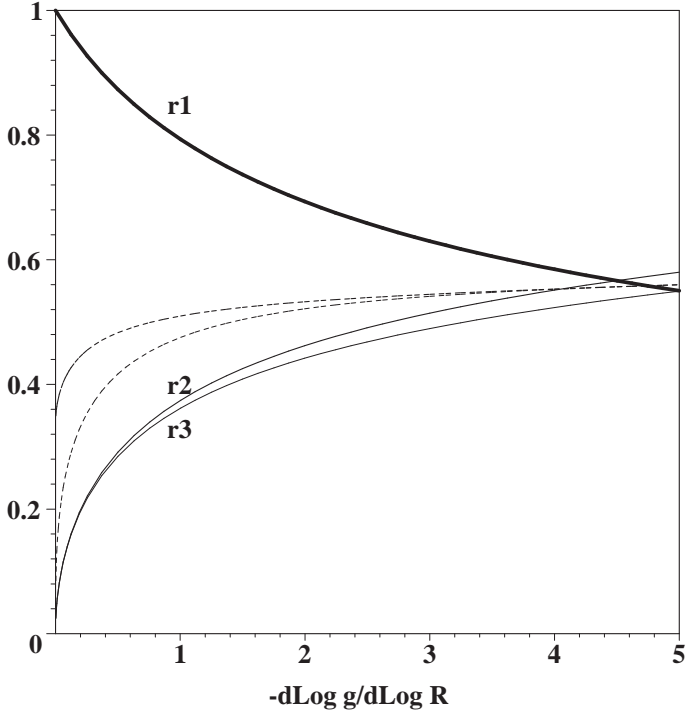}} 
\caption{
shows the Roche lobe semi-axis sizes $r_1, r_2, r_3$ of a satellite $m$ at orbital distance $R=D_o$ 
as functions of the gravity power-law index $\zeta = -{d \log g \over d \log D_o}$.  
Here the sizes are rescaled by $s \equiv \left({m \over M(D_o)}\right)^{1/3} D_o$.    Also shown are the arithematic mean 
$(r_1+r_2+r_3)/3$ (thicker dashed curve), and the geometric mean 
size $|r_1r_2r_3|^{1/3}$ (thin dashed curve).  }\label{semi}
\end{figure}

\begin{figure}{}
\resizebox{9cm}{!}{\includegraphics{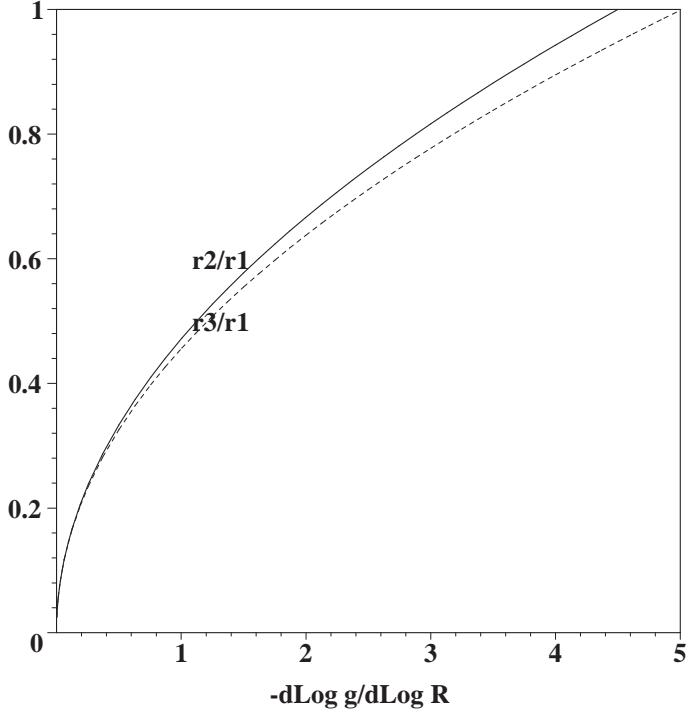}} 
\caption{
similar to the previous one, but shows the aspect ratio of the Roche lobe semi-axis sizes $r_2/r_1, r_3/r_1$ as functions of $\zeta = -{d \log g \over d \log R}$.  }\label{aspect}
\end{figure}

\section{Generalisation and Discussion}

\subsection{Extended spherical primary mass}

Although we refer to our primary mass as a mass point, our result applies exactly for a spatially extended spherical mass distribution of the primary $M(R)$, e.g., a galaxy, if we simply make the substitution
\beq
M \rightarrow M(R).
\eeq 
As a result the $\zeta-\Delta$ relation changes to 
\beq
2- {d\ln M \over d\ln R} = \zeta \Delta_1.
\eeq
As a specific example, a host galaxy can be approximated by a spherical Hernquist mass profile with
\beq
M(R)=M_0 \left({R \over R+ y_h R_0 }\right)^2, \qquad R_0 \equiv \sqrt{G M_0 \over a_0},
\eeq
with the scalelength parameter $y_h=1/2$, where we adopted notations of Zhao, Bacon, Taylor et al. (2005).
The Newtonian gravity of the model is
\beq
g_N = {G M(R) \over R^2} = a_0 \left(y_h+ y\right)^{-2}, \qquad y \equiv {R \over R_0}. 
\eeq
Note $g_N=a_0/(1+b)^2 = 4a_0/9 <a_0$ at $R=R_0$ and $g_N=a_0/b^2=4a_0$ is finite at $R=0$.
We give the Hernquist model a stronger-than-MOND gravity with $(k,n)=(9/10,1)$ 
so that the model has a rising terminal velocity curve.  
As shown by Fig.1b this model has $\zeta=0$ near the primary, and $\zeta=1/5$ far away. 

\begin{figure}{}
\resizebox{9cm}{!}
{\includegraphics{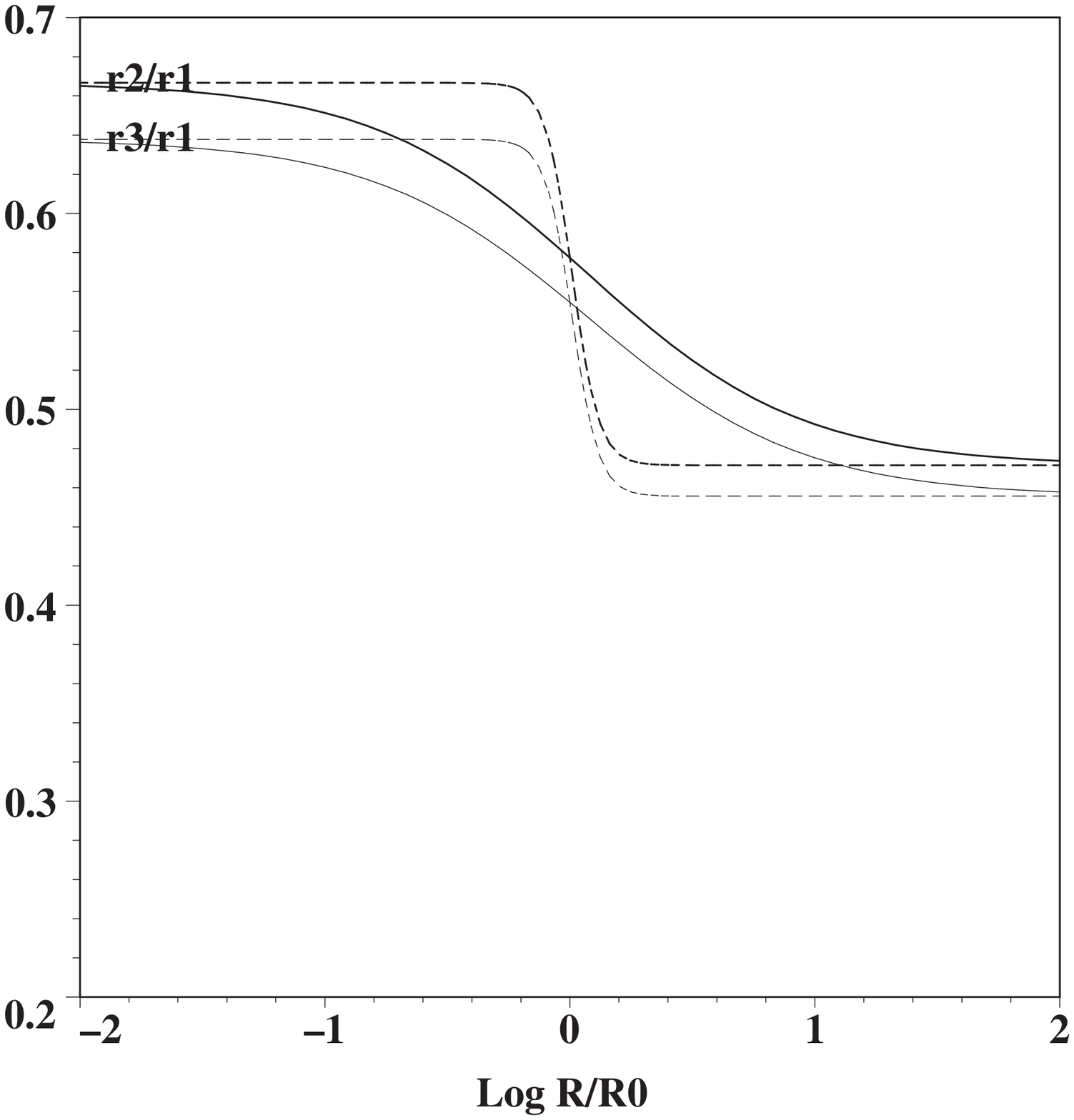},\includegraphics{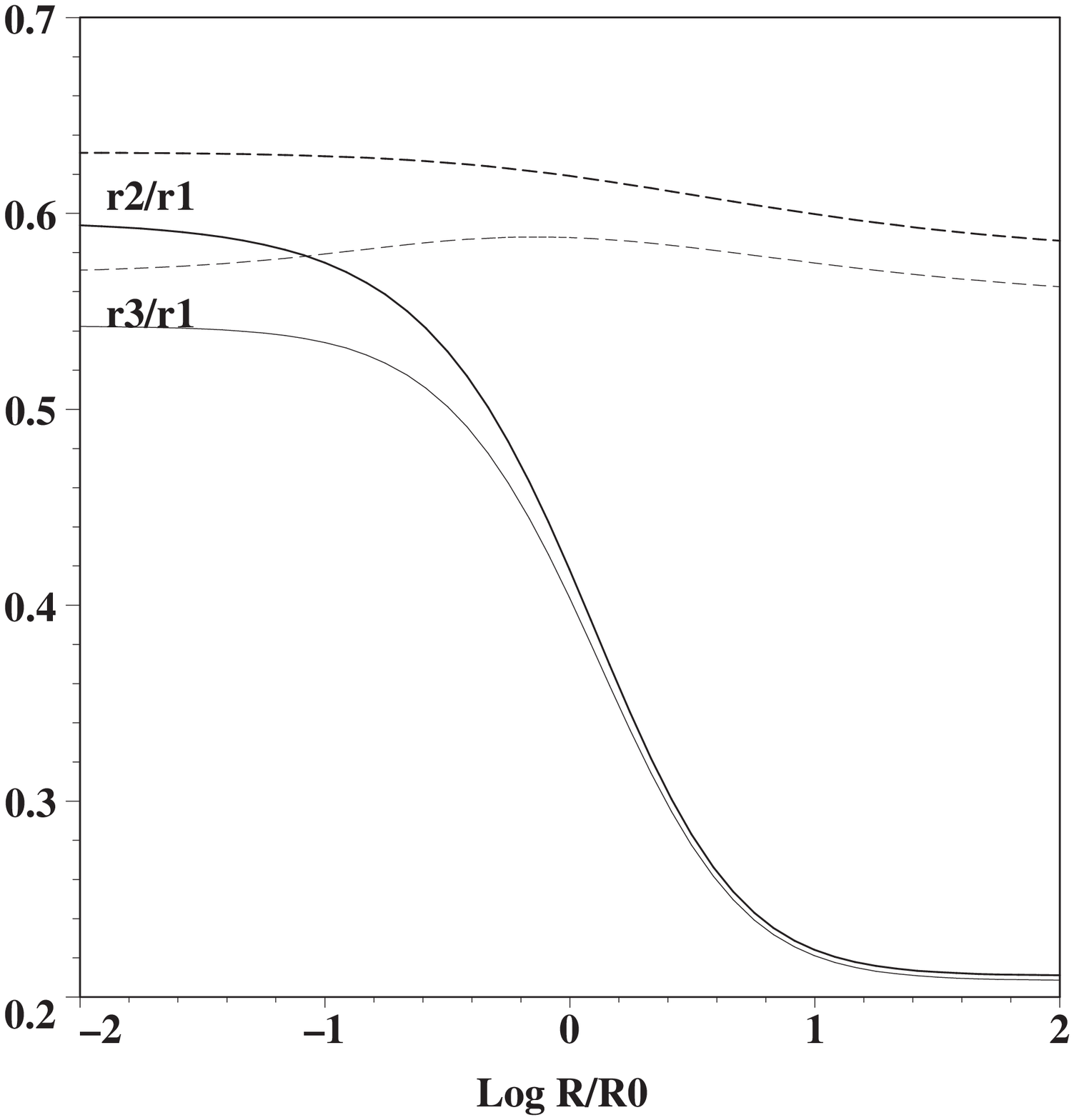}}
\caption{
showing the Roche lobe axis ratios $r_2/r_1, r_3/r_1$ as functions of $R/R_0$ with $R=D_o$ 
being the orbital distance of the secondary.  
{\bf Left panel}: Models are as in Fig.1a and Fig.1c, with $(k,n)=(1/2,1)$ (solid) and $(k,n)=(1/2,8)$ (dashed).  
{\bf Right panel}: as in Fig.1b and Fig.1d, with $(k,n)=(9/10,1)$ (solid) 
and $(k,n)=(1/4,1)$ (dashed) around a Hernquist profile mass. 
}\label{ratiomond}
\end{figure}

\subsection{Prolate or oblate?}

Interestingly, the second and the third (i.e., rotation) axes of the Roche lobe are always very similar in length.  This is 
because equation~(\ref{cubic}) contains a factor ${4 \over 27 \Delta_1\zeta_1} = {4 \over 27 (\Delta+3)} \ll 1$ for any 
$-1 < \Delta <\infty$.   In fact that 
\beq
{r_2 \over r_3} = [1, 1.02, 1.03, 1.045, 1.055, 1.065] \approx 1, 
\eeq
for $\zeta=[0,0.5,1,2,4,\infty]$, where $\zeta=1,2$ are the deep-MOND, and deep-Newtonian cases.

The first axis is often substantially longer than the other two, but not always.  We find for point masses
\beq
{r_1 \over r_2} = [\infty, 3, 2.12, 1.5, 0.95, 0], ~{\rm for~} \zeta=[0,0.5,1,2,5,\infty].
\eeq
The generally triaxial Roche lobe for point masses 
is nearly prolate for $0< \zeta <5$, and it becomes nearly spherical for $\zeta =4-6$,
and becomes nearly oblate only in exotic situations where the power-law index $\zeta=|d\ln g/d\ln R|$ is "harder" than 5.  
For extended mass, the ratio can be calculated from 
\beq
{r_1 \over r_2} = {3 \over 2} \sqrt{\Delta_1} ={3 \over 2}\sqrt{d\log g_N \over d\log g},
\eeq
together with a relation between $g$ and $g_N$ (cf. eq.~\ref{mu}) and $g_N={GM(R) \over R^2}$.

Overall the Roche lobe axis ratio changes by a significant amount for different assumptions of the 
laws of gravity and the mass profile (cf. Fig.~\ref{ratiomond}), e.g., 
the MONDian Roche lobe ($k=1/2$) at large distance is significantly more squashed than the Newtonian case, 
and a modified gravity with $k=9/10$ at large distance is even more squashed.  
We speculate that it might be feasible to differentiate these laws with a precise measurement of a Roche lobe.  
One advantage of using axis ratio is that it does {\it not} require precise knowledge of 
the satellite baryonic mass and its actual distance and absolute size; 
none of these factors enter equations~(\ref{ratio21}) and~(\ref{ratio32}).  In particular we note that
in Newtonian gravity $\Delta_1=1$, hence ${r_1 \over r_2} = {3 \over 2}$ holds for any mass distribution at any distance;
it would be interesting for testing non-Newtonian gravity to quantify the deviation from this ratio in observed Roche lobe.

\begin{figure}{}
\resizebox{9cm}{!}{\includegraphics{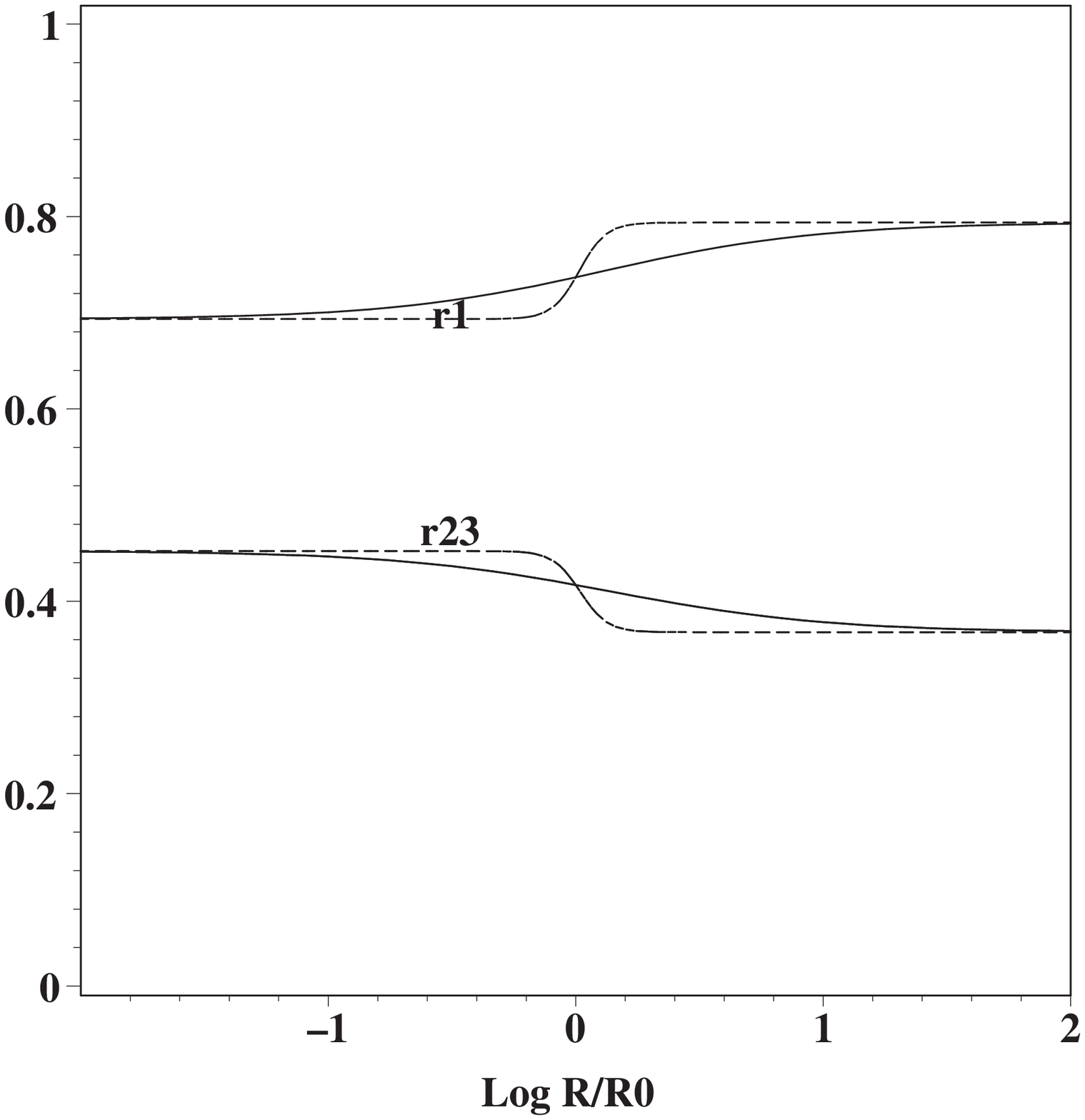},\includegraphics{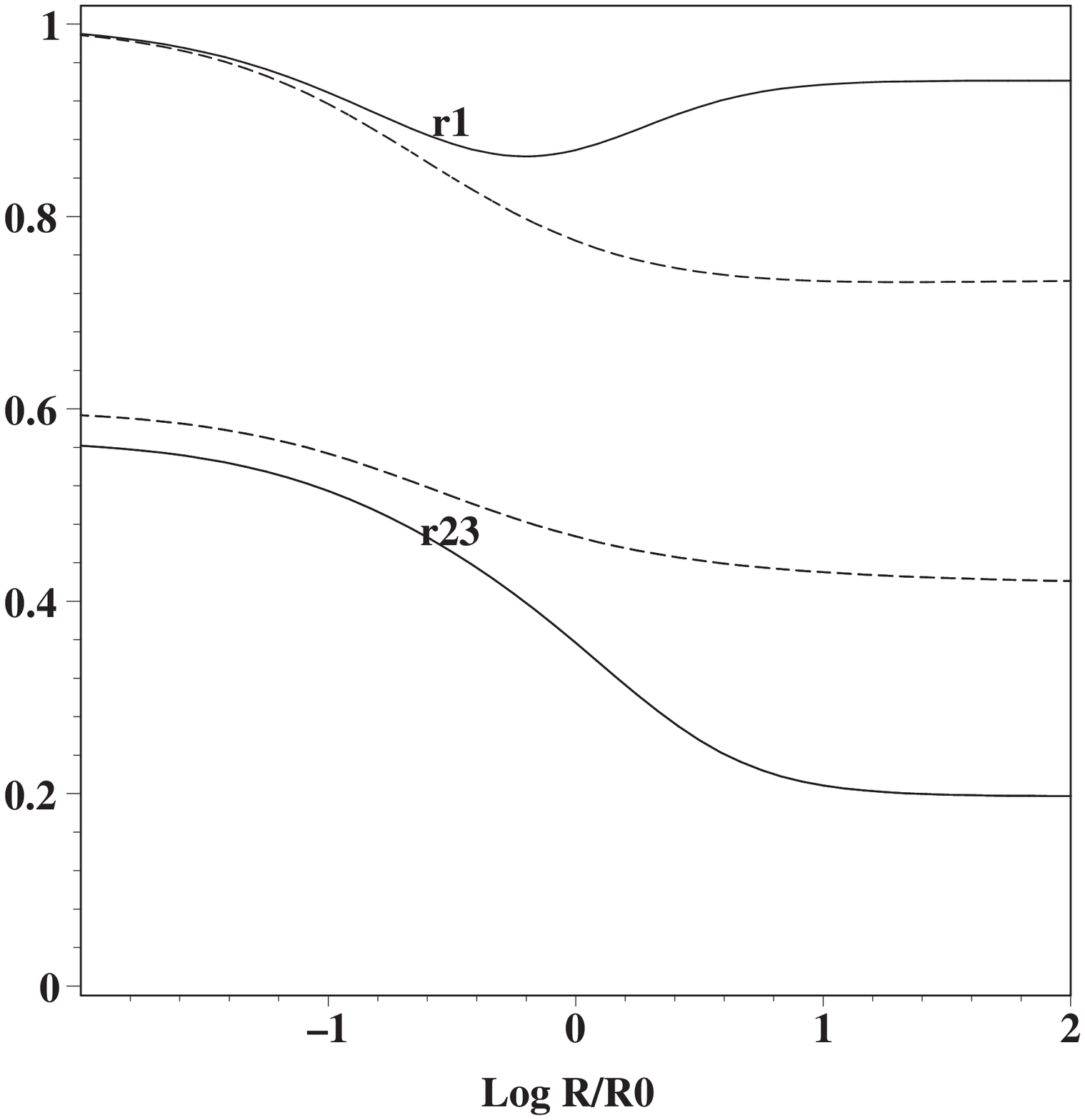}}
\caption{
showing the Roche lobe semi-axis sizes $r_1, \sqrt{r_2 r_3}$ as functions of $R/R_0$.  
Here the sizes are rescaled by $s \equiv \left({m \over M(R)}\right)^{1/3} R$ with $R=D_o$ 
being the orbital distance of the secondary.  {\bf Left panel}: models as in Fig.1a and Fig.1c, with $(k,n)=(1/2,1)$ (solid) 
and $(k,n)=(1/2,8)$ (dashed).  {\bf Right panel}: models as in Fig.1b and Fig.1d, with $(k,n)=(9/10,1)$ (solid) 
and $(k,n)=(1/4,1)$ (dashed) around a Hernquist profile mass. 
}\label{semimond}
\end{figure}

\subsection{Size-mass and size-distance relations}

In general, the Roche radii of two-body systems (whether stellar binaries or satellite-host galactic binaries) 
provide a measure of the relative weights of the binary in baryons in any MOND-like gravity.
It is interesting that the Roche Lobe size in any MOND-like gravity 
has the same simple scaling relation
\beq
{m \over r_1^3}/{M(D_o) \over D_o^3} = \zeta_1 = 2- {d\log V_{\rm cir}^2 \over d\log R}.
\eeq 
The above relation holds for an extended primary as well.
As in the Newtonian case the relevant mass is the unmodified true baryonic mass of the binary,
the difference is only in the prefactor, which depends on the $M$ and $D_o$.  
For point masses the prefactor varies from  $\zeta_1=3$ in the region with Keplerian rotation $V_{\rm cir}\propto R^{-1/2}$ 
(Binney \& Tremaine 1987) to $\zeta_1=2$ for the flat rotation curve region.  
A similar relation holds for $r_2$, 
\beq
\Theta_{\rm IR} \equiv 
{r_2 \over D_o} \left( { m \over M } \right)^{-1/3} = {2 \over 3 \Delta_1^{1 \over 2} \zeta_1^{1 \over 3} },
\eeq
where
\beq
\zeta_1 = 2- {d\log V_{\rm cir}^2 \over d\log R}, \qquad
\Delta_1 = {d\log g_N \over d\log g}.
\eeq
Here $\Theta_{\rm IR}$ takes the meaning of a rescaled Roche lobe intermediate semi-axis. 

The size of the lobe varies with the distance from the primary.
The Roche Lobe sizes in terms of $r_1$ and $\sqrt{r_2r_3} \sim r_2$ can be read off from Fig.~\ref{semimond} for different assumptions of the $\mu$-function and mass profile; in practise we can treat the Roche lobe as prolate.

\section{An application: Limiting radii of outer satellites of MW and M31}

As an illustration of the usuage of the Roche lobe, 
we compare the predicted Roche sizes with the observed limiting radii 
of outer satellites of galaxies.  Slightly different from Zhao (2005), 
the sample consists of all globular clusters and dwarf galaxies beyond 20 kpc (outside the disks) of both the Milky Way and Andromeda.   Only Andromeda satellites with reliable 3-dimensional 
distances to the Andromeda centre are used (McConnachie et al. 2004); tidal radius and distance of 
the newly discovered AndIX are from Harbeck et al. (2005).  
The limiting angular radii $\theta_{\rm lim,min}$ are taken primarily from Harris (1996) and Mateo (1998).  
Most systems are consistent with no flattening in projection.  For a few 
significant flattened systems we take the limiting radii on the {\it minor} axis.  
This radius should approximate the intermediate semi-axis $r_2$ of a nearly prolate Roche lobe if Roche lobes are filled;
a prolate object appears the same width on the minor axis at any inclination angle.  
Hence we have
\beq\label{fill}
{r_2 \over D_o} = F_1^{-1/3} {D \over D_\sun}{D_\sun \over D_o} \theta_{\rm lim,min}, \qquad F_1 \le 1,
\eeq   
where the factor $F_1 \le 1$ takes into the facts that (i) the Roche volume is smaller at
the pericenter of an eccentric orbit of a satellite, and that (ii) the volume filling factor of the 
pericentric Roche lobe can still be less than unity if star counts drop below detection limit before reaching
the edge of the lobe (e.g., Grebel et al. 2000).  Several distances are involved here,
$D_o$ is the present orbital distance, $D$ is the present distance of the Sun from the satellite, and $D_\sun$ 
is the Sun's distance to the centre of the host galaxy (here the Milky Way or the Andromeda).
The rhs is insensitive to errors on distance $D$; for the outer Milky Way satellites we have ${D \over D_o} \sim 1$,
and for Andromeda satellites we have ${D \over D_\sun} \sim 1$.   
Assuming mass-tracing-light, we expect that the satellite-to-host mass ratio is the ratio 
$q_*$ of observable stars within a factor of two, allowing for some differences in age and metallicity.
So we can define
\beq
F_2 \equiv {m \over M q_*} \sim 0.5-2.
\eeq 
Multiplying a factor $(m/M)^{-1/3}$ to both sides of the above equation~(\ref{fill}), we have 
\beq
{r_2 \over D_o} \left( { m \over M } \right)^{-1/3} \sim F_1^{-1/3} \Theta_{\rm min} \le \Theta_{\rm min}, 
\eeq
where\footnote{We drop the minor factor $F_2^{-1/3} \sim 0.8-1.25$.}
\beq
\Theta_{\rm min} \equiv {D \over D_\sun}{D_\sun \over D_o}  \theta_{\rm lim,min}q_*^{-1/3} , 
\eeq
is an observable, taking the meaning of the rescaled limiting minor axis size of a satellite.
If MOND-like mass-tracing-light models are correct, substitute in equations~(\ref{r1d})and~(\ref{ratio21})
we expect that 
\beq\label{ceiling}
\Theta_{\rm min} \ge \Theta_{\rm IR} = {2 \over 3 \Delta_1^{1 \over 2} \zeta_1^{1 \over 3} } \sim {2^{1/6} \over 3} = 0.374,
\eeq
where we used the fact that all the baryons are contained well inside 20kpc, and
\beq
\zeta_1 = 2- {d\log V_{\rm cir}^2 \over d\log R} \sim 2, \qquad
\Delta_1 = {d\log g_N \over d\log g} \sim 2,
\eeq
Note that we do not have freedom in chosing the law of the gravity, or $\zeta_1$ or $\Delta_1$:  
they are all determined by the nearly flat circular velocity curves in the Milky Way and Andromeda,
which are inferred from the fairly constant velocity dispersions of 
the outer satellites and other tracers of the outer halo such as PNe.

\begin{figure}{}
\resizebox{9cm}{!}{\includegraphics{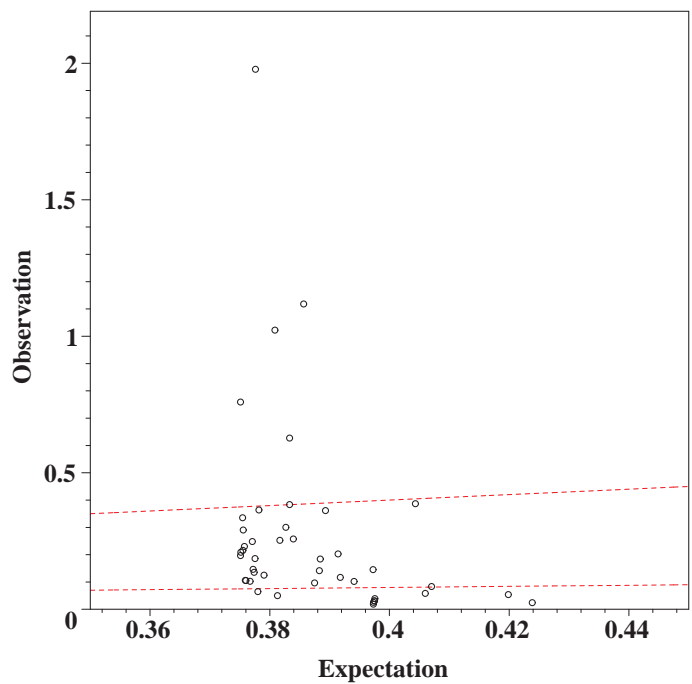}}
\caption{
shows expectations vs. observations of 
the rescaled limiting minor axis radii $\Theta_{\rm min}$ of outer satellites (beyond 20kpc) of the Milky Way and Andromeda.  The horizontal axis is the value of $\Theta_{\rm min}$ predicted from 
mass-tracing-light models assuming a satellite on a circular orbit filling its Roche lobe.
The expectation band is shown by two slanted dashed lines; 
the expectation-observation equality line (thick diagonal dashed line) 
should be an upper limit for satellites are on eccentric orbits and not filling their Roche lobes. 
There are several outliers.  To be conservative, we adopt 
a relatively low luminosity for the Milky Way and Andromeda so that 
$q_*={L_{\rm sat} \over 4\times 10^{10}\lsun}$.  
To be accurate we adopt a slightly falling circular velocity curve 
$V_{\rm cir}=V_0 \left[1+\left({r \over 100\kpc}\right)^2\right]^{-1/4}$ 
(cf. Wilkinson \& Evans 1999) for the Milky Way and Andromeda.  
}\label{compobs}
\end{figure}

Satellites with known proper motions often suggest only mildly eccentric orbits:
pericentric distances are rarely smaller than 1/2 of the present distances, and almost never 
smaller than 1/5 of the present distance (Dinescu et al. 1999, 2004 and references therein).
Outer globulars on long period orbits also have time to fill the instantaneous Roche lobe by
two-body relaxation (Bellazzini 2004). 
Globular clusters as a class generally exhibit extra-tidal stars under deep observations; 
a sharp break of the star count profile and/or sometimes two-dimensional confirmation of unvirialised structures
outside the King radius have been shown for globular clusters in the Milky Way, Andromeda, and other galaxies 
(Leon, Meylan \& Combes 2000, Lehmann \& Scholz 1997, Sohn et al. 2003, Harris et al. 2002). 
To be generous we assume the instantaneous Roche lobe has a filling factor $F_1 \sim 0.01-1$,
or $F^{-1/3}\sim F_1^{-1/3} \sim 1-5$.  So equation~(\ref{ceiling}) predicts that the majority of satellites
should lie in a band with dashed line boundaries as marked in Fig.~\ref{compobs}.

In reality we observe a number of outliers to the inequality equation~(\ref{ceiling}) and the expectation band.
Overall we observe a very large scatter of $\Theta_{\rm min}$ 
for satellites in similar gravitational field with similar expected values of $\Theta_{\rm IR}$ 
(cf. Fig.~\ref{compobs}).  Our result depends on the uncertain orbital pericentres of the halo satellites.
A better understanding of the issue will likely come 
with proper motions for a large sample of halo satellites from the GAIA astrometric mission.

\section{Conclusion}

The shapes of Roche lobes of a binary in a class of gravity theory including MOND are given analytically.  
The volume ratio of the Roche lobes is proven to scale linearly 
with the true mass ratio of the binary, hence models with dark matter
would predict a Roche lobe volume $V$ times that predicted in mass-tracing-light models, where the factor
$V=\left({M \over L}\right)_{\rm sat} \left({M \over L}\right)_{\rm host}^{-1}$ varies from $V \sim 1/10$ for typical 
outer globular clusters in a host galaxy to $V \sim 10$ for typical dwarf galaxies in a host galaxy.
Observed limiting radii of outer satellites of the Milky Way have a large scatter, with some outliers by a 
factor of 4 above or factor 10 below predictions of our basic models for the Roche lobes in MOND-like 
mass-tracing-light gravity theories.  
The flattenings of the Roche lobe are independent of the mass ratio, but are sensitive to the function
$\mu(g)$ in modifications to the law of gravity.  
Precise measurements of the exact shapes of other Roche-lobe-filling systems (e.g., mass-lossing stars or gas clouds 
in the outer halo of the Galaxy) might also test the law of gravity in the weak regime.

%\vfill\eject

\begin{acknowledgements} 
We thank the anonymous referee for helpful comments.
We acknowledge travel and publication support from Chinese NSF grant 10428308 to HSZ.
\end{acknowledgements}

%\vfill\eject

\end{document}